# Overview and design of the sPHENIX TPC

**H Klest[1]**

1 Stony Brook University, 100 Nicolls Rd., Stony Brook, NY 11790, USA

E-mail: Henry.Klest@Stonybrook.edu

**Abstract.** The proposed sPHENIX detector at RHIC is being designed to study with unprecedented precision the nature of the Quark Gluon Plasma, or QGP, at RHIC through measurement of jets, jet correlations, and Y particles. To reach the goal of producing a high mass-resolution reconstruction of the Y(nS) states, a compact TPC covering pseudorapidities < |1.1| is under construction to provide sPHENIX with the necessary invariant mass resolution. Similar to the ALICE-TPC upgrade, the sPHENIX TPC design presently utilizes quadruple-GEM detectors. As an alternative solution, a double-GEM-Micromega hybrid arrangement is currently being tested. Utilizing novel zigzag patterned readout pads, a modified SAMPA readout chip, and an optimized Ne-CF4 gas mixture immersed in a 1.4 T magnetic field to reduce ion backflow, sPHENIX will have a unique lens into the QGP. The motivation behind the design and innovations of the TPC will be presented, along with the current status of the project.

## 1. sPHENIX

The sPHENIX experiment is a state-of-the-art detector being designed for installation at the Relativistic Heavy-Ion Collider (RHIC) at Brookhaven National Laboratory. The physics goals of sPHENIX are measurements of jet substructure, fragmentation functions, heavy flavor, and individual Y(nS) states. A successor to the PHENIX experiment, sPHENIX will cover $0 \leq \phi \leq 2\pi$ in azimuth and pseudorapidities < |1.1|. sPHENIX will utilize the 1.4T superconducting solenoid used by the BaBar experiment at SLAC.

Three different tracking systems will allow sPHENIX to have good enough invariant mass resolution to measure individually the Y(nS) states. The most interior of these systems is the Monolithic-active-pixel-sensor-based Vertex Detector or MVTX. The MVTX is comprised of three layers of silicon pixel detectors extending from 2.3 cm < r < 3.9 cm around the interaction point and will provide high-resolution point measurements of particle locations for use in vertexing [1]. The acceptance around the interaction point will be z < 10 cm and $0 \leq \phi \leq 2\pi$. At a higher radius than the MVTX detector will be the Intermediate Tracker or INTT extending from 6 cm < r < 12cm. The INTT is a silicon strip detector for pileup event separation and additional tracking. The final tracking element, spanning from 20 cm < r < 78 cm and z < |1.05m|, is the compact time-projection chamber.

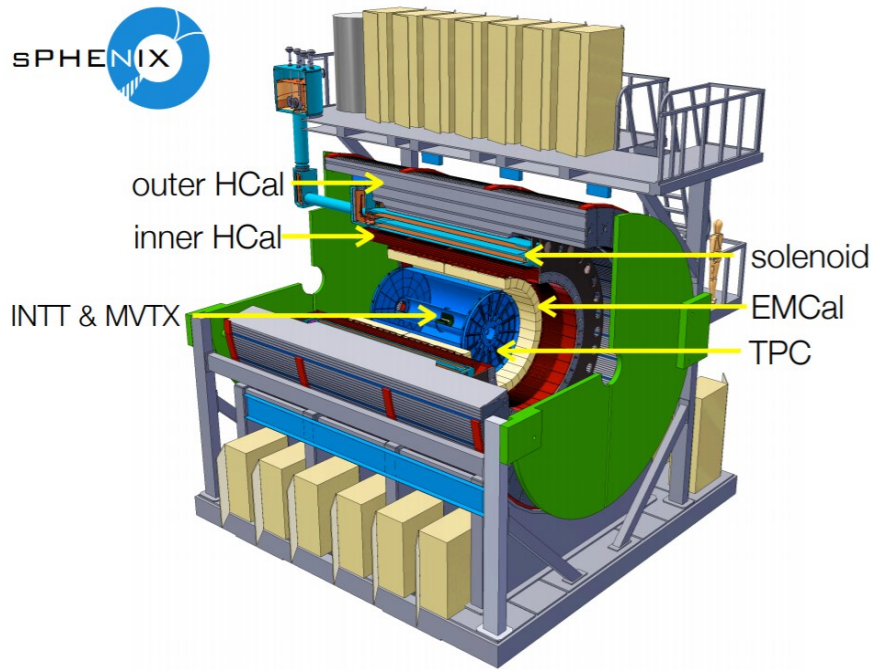

**Figure 1**. Proposed sPHENIX Design [2]

Outside the tracking systems, in order of increasing radius, will be an electromagnetic calorimeter, an "inner" hadronic calorimeter, the solenoid, an "outer" hadronic calorimeter, and a support structure.

**2. sPHENIX TPC**

The TPC will be the main tracking element for sPHENIX. It will produce a targeted invariant mass resolution of less than 100 MeV in the di-electron decay channel of the Y, which is the anticipated invariant mass resolution needed to resolve the Y(nS) states from one another, as can be extrapolated from simulation data in figure 2.

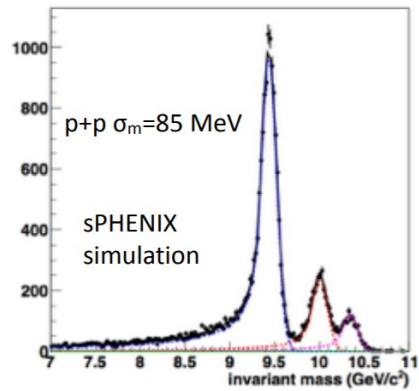

**Figure 2.** sPHENIX Simulation of Y(1s+2s+3s) yield for p+p Collisions

The TPC will also have good invariant mass resolution down to transverse momenta on the order of hundreds of MeV. It is well known that the energy loss of partons inside the QGP can be probed by observing the suppression and angular correlation of mesons containing charm and bottom quarks. These measurements are enabled by the displaced vertexing ability of the MVTX in conjunction with the low-$p_t$ tracking capabilities of the TPC [3].

*2.1. Design*
The sPHENIX TPC is compact; its small coverage in radius will make it almost one third the outer radius of the ALICE TPC, and one half the outer radius of the STAR TPC. The sPHENIX TPC will be filled with a mixture of neon and $CF_4$, which will be ionized when charged particles pass through it. A central membrane parallel to the endplates will hold a negative voltage on the order of tens of kV that will accelerate the ionized electrons towards the ends of the cylinder. An inner and an outer field cage made of copper circuit cards will keep the electric field uniform in the |z|-direction throughout the TPC by carrying successively decreasing voltages in z.

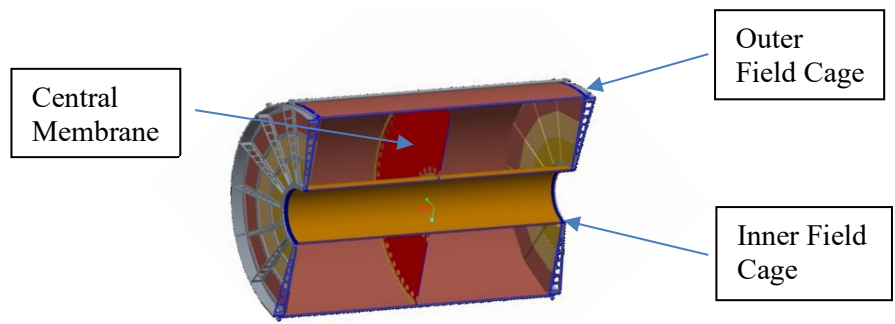

**Figure 3.** The sPHENIX TPC Mechanical Design

Following the ALICE, STAR, and ILC TPC designs, the sPHENIX TPC (figure 3) will also utilize micropattern gas detectors (MPGDs) on the ends of the cylinder for amplification and detection of the ionized electrons. MPGDs provide cost-effectiveness, good spatial resolution, and high rate capability. The instrumented region of the TPC is divided into 72 separate arc-shaped readout sections utilizing MPGDs, segmented 3 times in r and 24 in ϕ.

*2.1.1 Spatial Resolution.* In order to achieve an invariant resolution less than 100 MeV, the TPC will need to attain an intrinsic spatial resolution around 100 μm. MPGDs such as gas electron multipliers (GEMs) [4] and micromesh gaseous structures (Micromegas) [5] are known to be capable of producing this resolution, but the major limiting factor to the spatial resolution for sPHENIX is the buildup of ions in certain regions of the TPC causing distortions of the electric field. For sPHENIX to achieve the desired resolution, the percentage of ions that return to the drift region from the MPGD must be less than 1% when operated in a continuous, ungated fashion. The electrons do not affect this buildup because they are lighter than the ions and are easily swept out of the gas and into the gain stage. Such a buildup of ions in a TPC drift volume is known as space charge.

To some degree, space charge is unavoidable in a gas TPC because the operation requires the production of ions. However, in the amplification stage of most MPGDs, a significant number of ions are produced. To minimize space charge, these ions should be prevented from flowing back into the drift region of the TPC [6]. Both GEMs and Micromegas have the unique property of stopping most ions from returning to the drift region by capturing them. A quad-GEM (figure 4) and a two-GEM-Micromegas (figure 5) configuration are both being investigated as possible options for use as the gain stage of the TPC.

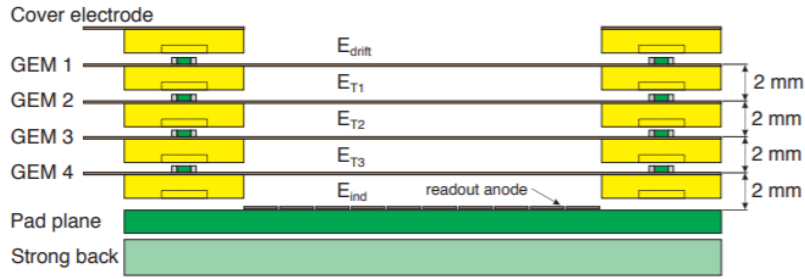

**Figure 4.** Quad-GEM Configuration from ALICE TDR [7]

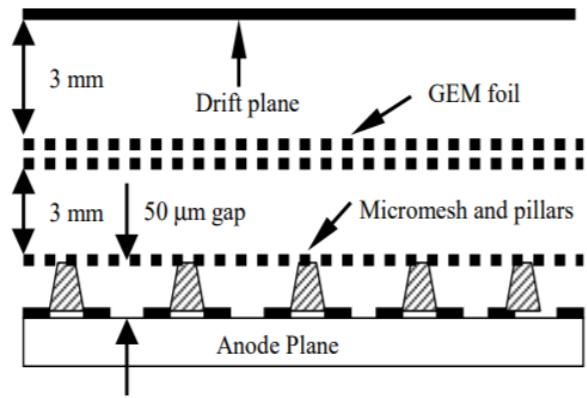

**Figure 5.** Two-GEM-Micromegas Configuration [8]

The compact TPC could utilize a quadruple GEM voltage scheme devised by ALICE and nominally achieve an IBF percentage of 0.15%.

An attractive option for reducing IBF is a passive gating grid. A grid or a series of wires at alternating values of potential held suspended in a magnetic field are capable of being transparent to most electrons while being opaque to ions as shown in figure 6.

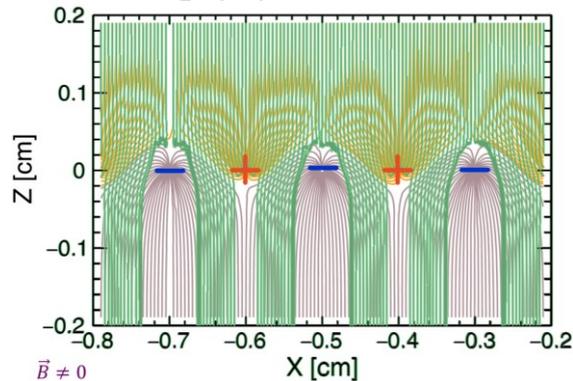

**Figure 6.** Simulation of Electron (Green) and Ion (Orange and Purple) Trajectories

The electron-transparency and ion-opacity of such a grid is a result of the well documented E x B effect and the large mass difference between the electrons and ions. The mechanics of holding the wires at set potentials while also keeping them at set distances from each other is currently undergoing R&D.

Another method of mitigating space charge is to only readout the portion of the TPC where space charge is the least problematic. The field cages attract space charge; due to the lesser radius, the inner field cage will have the highest density of space charge distorting the electric field. As a result, only the region of 32cm < r < 78 cm will be instrumented, which will cause the space charge that accumulates on the inner radius of the TPC to minimally affect the spatial resolution of the instrumented region.

Space charge is not the only factor affecting spatial resolution. The intrinsic spatial resolution is also a function of readout pad shape and size. Almost all MPGDs utilize copper pads underneath the gain stage to pick up the electron signal produced by amplification. The standard pad shape is a rectangle, as rectangles allow for simple geometric calculations to reproduce the location of the electron cloud. For the sPHENIX TPC, the readout pads will be zigzag shaped and average 2mm x 12.5mm in size. The distribution of charge on the pads from a simulated event is shown in figure 7.

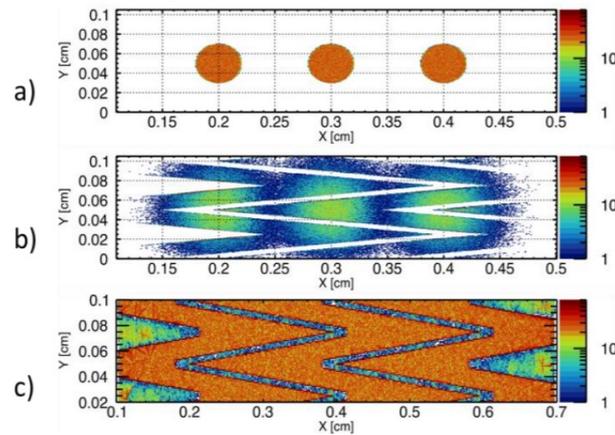

**Figure 7.** Simulation of Zigzag Segmented Pads, A) Three clouds of charge dropped from 2mm above pads. B) Charge density at pad level. C) Charge distribution resulting from a plane of uniform charge [8].

This zigzag segmentation allows for greater charge sharing between pads, which enables better reconstruction of the location of charge clouds incident on the pad plane. This also means that larger pads can be used, thereby decreasing channel count. It is extremely unlikely that only a single zigzag pad will be hit. However, the possibility of a single rectangular shaped pad being hit imposes a fundamental limit on the spatial resolution of rectangular pads. As illustrated in figure 8.

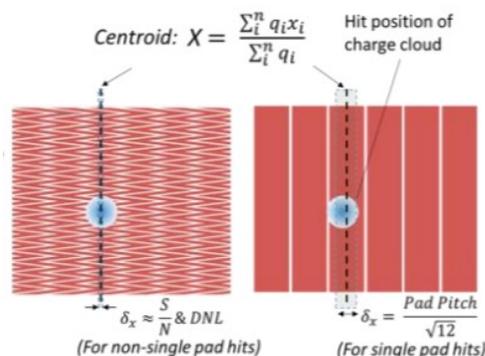

**Figure 8.** Visualization of the Advantage of Zigzag-shaped Pads [9].

Previous studies of zigzag patterned pads examined different geometries that were susceptible to poor resolution when rotated. However, the geometry of zigzag-shaped pads to be used in sPHENIX is effectively rotationally invariant with respect to spatial resolution, since the centroid of the charge cloud can be calculated with equal effectiveness regardless of rotation.

*2.1.2 Rate Capability.* Since RHIC will have an event rate between 50 and 200 kHz by the time sPHENIX is completed, the TPC will be required to handle high multiplicities of charged particles passing through it at any given time. By virtue of the tunability of gas mixtures and voltages, MPGDs provide an attractive option for use in high-rate applications. An optimized mixture of Ne-$CF_4$, a gas with a high drift velocity, will allow the TPC to feasibly handle an event rate of 50 kHz or higher. This high event rate also increases the rate of space charge creation in the TPC. A laser calibration system with a high repetition rate is currently being designed to allow for active correction of tracks that have been distorted by space charge. The laser will mimic particle tracks by ionizing the gas in a known pattern. The data acquired can then be compared to the known laser pattern and corrected for distortions.

The type of electronics used for detection of electrons in the instrumented region is an important factor that contributes to both readout rate and spatial resolution. An overview is provided in figure 9.

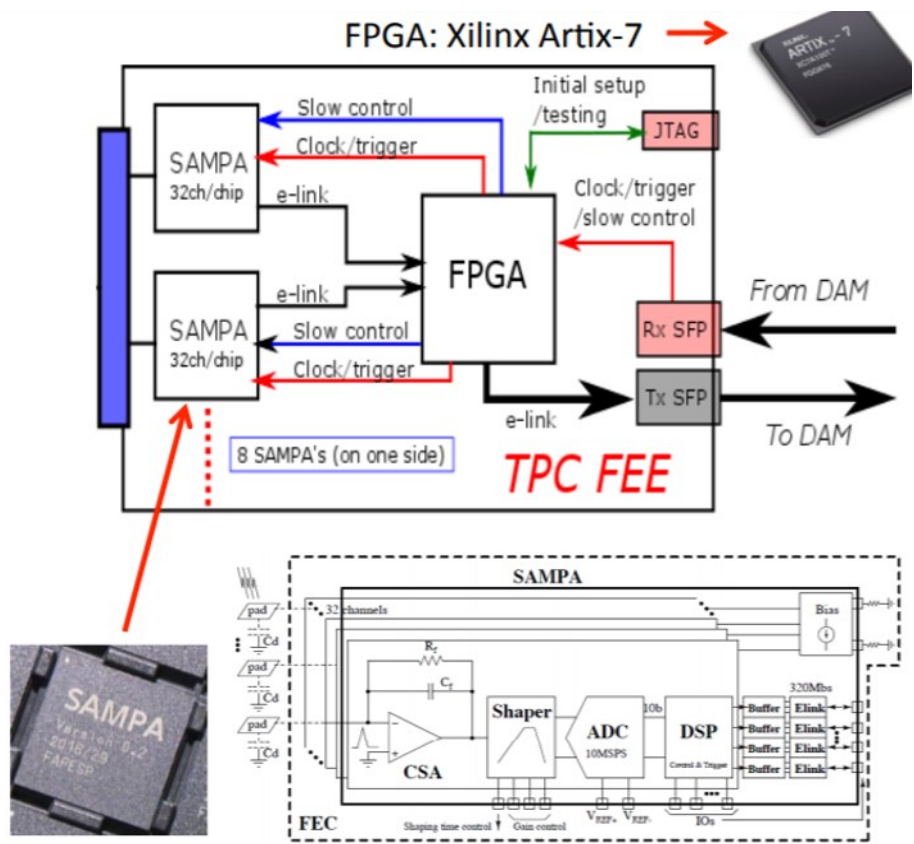

**Figure 9.** sPHENIX Electronics Overview

The sPHENIX TPC will be read out in a continuous and gateless fashion that will require the SAMPA to define event boundaries offline. The SAMPA chip is capable of handling 32 readout channels with an 80 ns peaking time [10]. This fast timing is critical to sPHENIX, as the high event rate and fast drift velocity mean that individual signals will be spaced more closely in time than those of ALICE, which utilized the 160 ns peaking time option of the SAMPA. Software further down the data stream will disentangle stacked events. The planned sPHENIX datastream is outlined in figure 10.

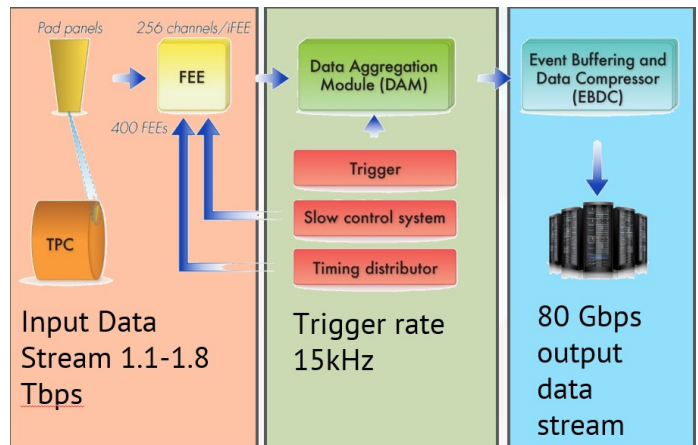

**Figure 10.** sPHENIX Datastream Overview

*2.1.3 Size Constraints.* The sPHENIX compact TPC merges the gas volume barrier and the field cage in order to save space for other detector elements inside the solenoid. The solenoid has a radius of 1.3 meters and will contain the MVTX, INTT, and TPC, in addition to the inner electromagnetic and hadronic calorimeters as shown in figure 11.

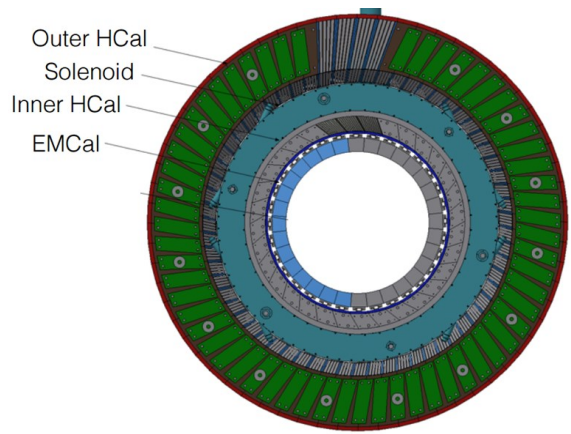

**Figure 11.** sPHENIX Calorimetry Layout

A primary motivation for the compactness of the TPC is the uniformity of the magnetic field produced by the solenoid. Since a uniform magnetic field is an important part of the TPC's principle of operation, the TPC should make use of a portion of the solenoid interior with a relatively uniform field.

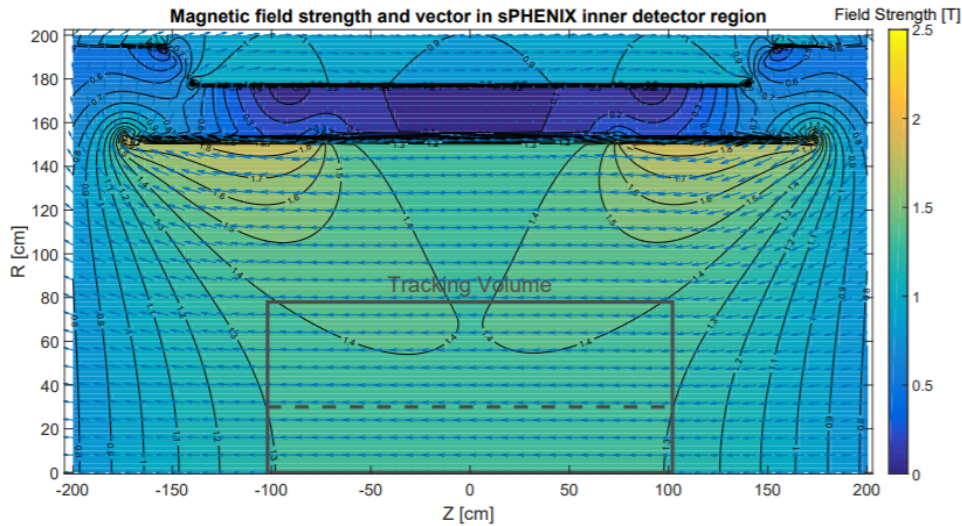

**Figure 12.** Field Map of the sPHENIX Solenoid

The calorimeters outside the TPC are free to occupy the region with a less uniform magnetic field, as calorimetry generally does not require a uniform magnetic field. Since the region nearest to the solenoid has the greatest field distortions and the sPHENIX jet measurements rely on calorimetry, calorimeters provide a good option to fill this space. As can be seen in figure 12, the TPC resides in a section of the solenoid that has a uniform magnetic field in z. As a result, the tracking capabilities of the compact TPC will be minimally affected by distortions in the magnetic field that would otherwise be significant.

## 3. Summary

sPHENIX and the sPHENIX TPC design considerations have been briefly summarized in these proceedings. The sPHENIX TPC is on track to attain the required invariant mass resolution for performing Y(nS) spectroscopy as well as enabling measurements of jet substructure, fragmentation functions, and parton energy loss inside the QGP.